\documentclass[twocolumn,showpacs,floatfix]{revtex4}
\usepackage{bm}
\usepackage{amssymb}

\usepackage{graphicx}
\usepackage[center]{subfigure}

\begin{document}

\title{Renormalization group theory for the phase field crystal equation}

\author{Badrinarayan P. Athreya$^1$, Nigel Goldenfeld$^2$, and Jonathan A. Dantzig$^1$}
\affiliation{$^1$Department of Mechanical and Industrial
Engineering,\\
University of Illinois at Urbana-Champaign,\\
1206 West Green Street, \\
Urbana, IL 61801\\
\\
$^2$Department of Physics,\\
University of Illinois at Urbana-Champaign,\\
Loomis Laboratory, 1110 West Green Street, \\
Urbana, IL 61801.}

\begin{abstract}
We derive a set of rotationally covariant amplitude equations for use
in multiscale simulation of the two dimensional phase field crystal
(PFC) model by a variety of renormalization group (RG) methods.  We
show that the presence of a conservation law introduces an ambiguity in
operator ordering in the RG procedure, which we show how to resolve. We
compare our analysis with standard multiple scales techniques, where
identical results can be obtained with greater labor, by going to sixth
order in perturbation theory, and by assuming the correct scaling of
space and time.
\end{abstract}

%

 \maketitle

\section{Introduction}
A fundamental theoretical and computational challenge in materials
modeling is that of simultaneously capturing dynamics occurring over a
wide range of length and time scales, under processing conditions. A
classic example of such a multiscale problem is dendritic growth, a
phenomenon seen in the solidification of undercooled melts, which
involves the capillary length $\sim10^{-9}$m, the scale of the pattern
$\sim10^{-6}$m, and the diffusion length $\sim10^{-4}$m, which together
span length scales over five orders of magnitude, and heat/solute
transport through diffusion which occurs on time scales of
$\sim10^{-3}$s. Only recently, after considerable advances in computing
technology, and through the use of sophisticated computational
techniques \cite{PGD,Jeong2,Jeong1}, has this problem become tractable
in three dimensions. Although a number of computational
approaches\cite{Phillipsbook, VVED04} including quasi-continuum
methods\cite{Tadmor, Shenoy, Ortiz, Miller}, the heterogeneous
multiscale method\cite{Weinan1, Weinan2}, multi-scale molecular
dynamics\cite{Rudd, Kaxiras, Robbins, CURT02}, multigrid
variants\cite{Fish} and extensions of the phase field
model\cite{warren03} have been proposed, they all appear to have
significant limitations. Developing a ``handshake'' algorithm, to
seamlessly integrate the transition between scales, is a common
problem. Also, most models appear to be capable of handling only a
limited number of crystallographic orientations, with a few isolated
defects.

Elder \emph{et al.} \cite{ekhg02,eg04} recently proposed a continuum,
non-linear partial differential equation, which they called the phase
field crystal (PFC) model, for realistically describing materials
processing phenomena in polycrystalline materials. The PFC equation
describes the evolution of the time averaged density field in the
material, subjected to the essential constraint of mass conservation.
While averaging the density makes it possible for the model to capture
phenomena over diffusive time scales (not possible with molecular
dynamics), the spatially periodic variations, with a wavelength on the
order of the inter-atomic spacing, allow it to incorporate lattice
defects such as vacancies. These rapid spatial variations, however, are
also the bane of the model, as they necessitate the use of a uniformly
fine computational mesh whose grid spacing approaches the interatomic
separation. Thus the PFC model is computationally expensive for
mesoscale problems (such as dendritic growth), although still
considerably better than a molecular dynamics calculation.

We have recently described a theoretical approach to this difficulty
\cite{GAD05_1, GAD05_2}, presenting a heuristic renormalization group
(RG) \cite{NGbook} method to coarse-grain the PFC equation and obtain
equations of motion for the amplitude and phase of the periodic density
field. Using these variables, it is possible to reconstruct the
original field to a certain order of approximation. The main advantage
of such a description is that the amplitude and phase of the density
field vary on length scales much larger in comparison to those of the
field itself, which enables us to use a coarser mesh to speed up
calculations (see \cite{GAD05_1} for details on accuracy and speedup).
Furthermore, the relative uniformity of these variables permits
solution of the equations governing them on an adaptive grid which, we
anticipate, will further improve computational efficiency.  We
emphasise that the RG procedure is more than a naive coarse-graining in
real space of the density, because it uses the dynamics inherent in the
underlying equation to project out the long-wavelength, small-frequency
behavior.  The essential effect of this distinction is that valuable
aspects of the phase field crystal model, such as its native inclusion
of elasticity, are preserved. In \cite{GAD05_1, GAD05_2} we verified
explicitly, by numerical calculation, how the RG equations are able to
accomplish this, and thus are suitable for dealing with polycrystalline
systems.

The main purpose of the present article is to present full details of
the systematic calculation to derive such coarse-grained equations from
the PFC equation. A secondary goal is to compare and contrast the
variety of techniques that are available to derive coarse-grained
equations of motion. A rather surprising finding of our work was that
when we followed naively the \lq\lq cookbook recipe" for each method,
our results were not identical, with the RG methods yielding a form of
the amplitude equation, slightly different from that derived by the
classical method of multiple scales. The PFC equation obeys a local
conservation law, and while this by itself can lead to a variety of
interesting features\cite{MATT00}, it also brings to the fore an
ambiguity with the usual implementation of the renormalization
procedure, something that is not unique to conservation laws.  This
ambiguity is essentially an operator ordering one, and can be remedied
in a straightforward way. Once done, all methods yield the same
amplitude equation, even though the technical details are quite
distinct in the different methods. As a pedagogical exercise, we
present the analysis for the Van der Pol oscillator in the Appendix,
once again obtaining consistent results from all methods when using the
approach described herein. Our main conclusion is that the
renormalization group method is still considerably easier to implement
than competing approaches, and in particular requires no knowledge of
the scaling relationship between space and time while achieving full
rotational covariance at lowest order in perturbation theory.

In the remainder of this Introduction, we review the main conceptual
developments leading up to the techniques described in this paper.  It
is now relatively standard in non-linear pattern formation problems to
use amplitude equations to uncover universal features of pattern
forming systems. The formalism, first introduced by Newell, Whitehead,
and Segel (NWS) \cite{New_Whit69,Segel69} to describe periodic patterns
in Rayleigh-B\'{e}nard convection, offers a way to extract the
spatio-temporal envelope of these patterns, which then allows one to
predict the dynamics qualitatively with very little information about
microscopic details. Unfortunately however, the NWS equation, as
originally constructed, can only describe the dynamics of patterns
oriented along the same fixed direction, everywhere in space, whereas
physical systems often produce complex mosaics of patterns with no
particular orientational preference. Such mosaics arise in real systems
which are invariant under rotations, and hence any equation which is
used to study them must also have the crucial property of rotational
covariance, something that is lacking in the NWS equation. Equations
with an orientational bias can be very difficult to implement
numerically, especially on systems with arbitrarily oriented patterns.
Nevertheless, the NWS equation embodies the important notion of
coarse-graining, which has played a significant role in shaping the
modern day theory of pattern formation \cite{cross}, and also forms the
basis of our approach to multiscale modeling with the PFC equation.

Gunaratne \emph{et al.} \cite{gunaratne} first derived a rotationally
covariant form of the NWS equation using the method of multiple scales
\cite{Sturrock57,Frieman63,Orszagbook,Nayfehbook}, where they assumed
isotropic scaling of the spatial variables. They showed that the
spatial operator in the NWS equation could be \emph{symmetrized}, by
systematically extending the calculation to higher orders in the
perturbation parameter $\epsilon$, the reduced Rayleigh number. They
explained that the finite truncation of the perturbation series
destroyed the rotational symmetry of the operator, which could however
be recovered at a higher order. Another important conclusion of their
work was that the qualitative behavior of pattern formation remained
unchanged if one ignored higher order corrections, provided the
equation itself was rotationally covariant. A drawback of their
calculation, however, was (as with any application of the method of
multiple scales), the need to guess {\it a priori\/}, the correct
scaling of space-time variables. In addition, their calculation
required gradual accumulation of operators and terms up to
$\mathcal{O}(\epsilon^4)$, before a rotationally covariant equation
emerged.

A more systematic approach emerged shortly after: Chen \emph{et al.}
showed how to perform reductive perturbation theory using RG
methods\cite{CGO2}, and obtained the NWS equation for the
Swift-Hohenberg equation\cite{SH} by renormalizing the leading secular
divergences at each order. Graham\cite{Graham} subsequently showed
that, in fact, this method gave the fully rotational covariant
equations, if all secular terms are renormalized and a careful choice
of operator splitting is used. Calculations involving the RG typically
produce elegant and accurate uniformly valid approximations for
ordinary differential equations (ODEs), starting from simple
perturbation series where no knowledge of the scaling present in the
system is exercised\cite{CGO2}.  For partial differential equations
(PDEs), the same approach is successful, but generates a tedious number
of perturbation terms at higher orders. This difficulty arises from the
need to explicitly construct secular solutions of the highest possible
order, at every order in $\epsilon$, and from a practical standpoint
equals (if not outweighs) the advantage of requiring no prior insight
into the problem. For this reason, calculations employing this method
for PDEs have rarely gone beyond $\mathcal{O}(\epsilon)$. The key
advantage of the RG method, however, is that when carefully performed,
the calculation yields a rotationally covariant amplitude equation at a
much lower order in $\epsilon$ compared to the method of multiple
scales, as was shown by Graham \cite{Graham}.

Nozaki \emph{et al.} \cite{Nozaki00,Nozaki01}, have developed a more
abstract version of the perturbative RG for weakly non-linear PDEs,
called the ``proto-RG'' scheme. They argue that if one is willing to
sacrifice some of the purely mechanical aspects of the conventional RG
by taking cognizance of the system's properties, such as those
exhibited by the governing differential equation, one can obtain a
rotationally covariant amplitude equation to $\mathcal{O}(\epsilon)$
without having to construct any secular solutions. By computing minimal
particular solutions, usually obtained by a straight-forward
inspection, one can even obtain $\mathcal{O}(\epsilon^2)$ corrections
with only a little more algebra. They illustrated the relative
simplicity of this method by deriving the rotationally covariant form
of the NWS equation to $\mathcal{O}(\epsilon^2)$, as previously derived
by Gunaratne \emph{et al.} \cite{gunaratne}. Shiwa \cite{Shiwa} further
demonstrated the efficacy of this scheme by obtaining the well-known
Cross-Newell phase equation \cite{CN,PN}, which describes phase
dynamics of patterns generated by the Swift-Hohenberg equation. A
drawback of this approach is in the selection of the so called
``proto-RG'' operator, which turns out to be non-unique in general.
Nozaki \emph{et al.} show, however, that the operator \emph{is}
uniquely specified, provided we insist on the lowest order differential
operator possible.

As the reader will have no doubt realized from this synopsis, several
methods and variants exist for deriving envelope equations from order
parameter equations (OPEs) that produce predominantly periodic
patterns. Although one may argue that some of these methods are
essentially variants of perturbative RG theory for PDEs, they are
structurally very different. It is thus a very instructive exercise to
compare the defining properties of each of these methods in the context
of a single microscopic OPE. We present such a detailed study in this
paper using the PFC equation, which unlike the Swift-Hohenberg model,
has not been extensively studied.

The article is organized as follows. To set the context for our work,
we briefly introduce the PFC model in section \ref{PFC}. In section
\ref{QD} we present a detailed derivation of an amplitude equation from
the PFC model using a heuristic approach. In section \ref{PRG}, we use
the proto-RG method to derive the amplitude equation more
systematically. We attempt to verify these calculations independently
in section \ref{MS} using the method of multiple scales. A 1-D
derivation via the conventional RG method is presented for completeness
in section \ref{RG}. We find that while the proto-RG and RG results are
consistent, they do not agree with the other calculations, due to an
operator ordering ambiguity not previously noticed. We remedy this in
\ref{MODI}, and conclude with some remarks in section \ref{CONC}.

\section{\label{PFC}The phase field crystal equation}

The phase field crystal model proposes a sixth order non-linear partial
differential equation for describing the space-time evolution of the
time-averaged, conserved density variable $\psi(\mathbf{x},t)$, of a
material. As has been shown \cite{ekhg02,eg04}, this equation has the
potential to accurately model a variety of key materials processing
phenomena, including heterogeneous nucleation and grain growth, liquid
phase epitaxial growth, ductile fracture mechanics, dislocation
mechanics \cite{BGE05}, and plasticity. An important feature which
differentiates it from another popular continuum material model, the
phase field model \cite{warren03}, is its incorporation of elasticity
in the free energy functional through terms that guarantee gradients in
the equilibrium density field, for certain values of the control
parameters. Crucial to the construction of this free energy, is the
observation that elasticity is a natural property of a system which is
characterized by periodic fields. We refer the reader to the exhaustive
article by Elder and Grant \cite{eg04} for a detailed description of
the model and its applications.

From the point of view of pattern formation theory, the PFC equation is
the conserved analog of the simplest form of the Swift-Hohenberg
equation (with only the cubic non-linearity), and is written as
\begin{equation}
\label{eqn:pfc} \partial_t \psi = \nabla^2\left[\left\{(1 + \nabla^2)^2 - \epsilon\right\} \psi
+\psi^3\right]+\zeta.
\end{equation}
Here, $\epsilon$ is the scaled undercooling, a parameter akin to the
modified Rayleigh number, controlling the stability of the uniform
phase $\bar{\psi}$ (liquid) to the appearance of either a periodic
striped phase or a periodic hexagonal phase (both crystalline solids),
and $\zeta$ is the conserved Gaussian noise which accounts for thermal
fluctuations in the system. A phase-diagram illustrating the phase
boundaries and coexistence curves in $\epsilon-\bar{\psi}$ space is
given in \cite{eg04}.
\begin{figure}[htbp]
\begin{center}
\subfigure[\label{fig:pfc1} $t=300$] {\includegraphics[height=2.5in,angle=0]{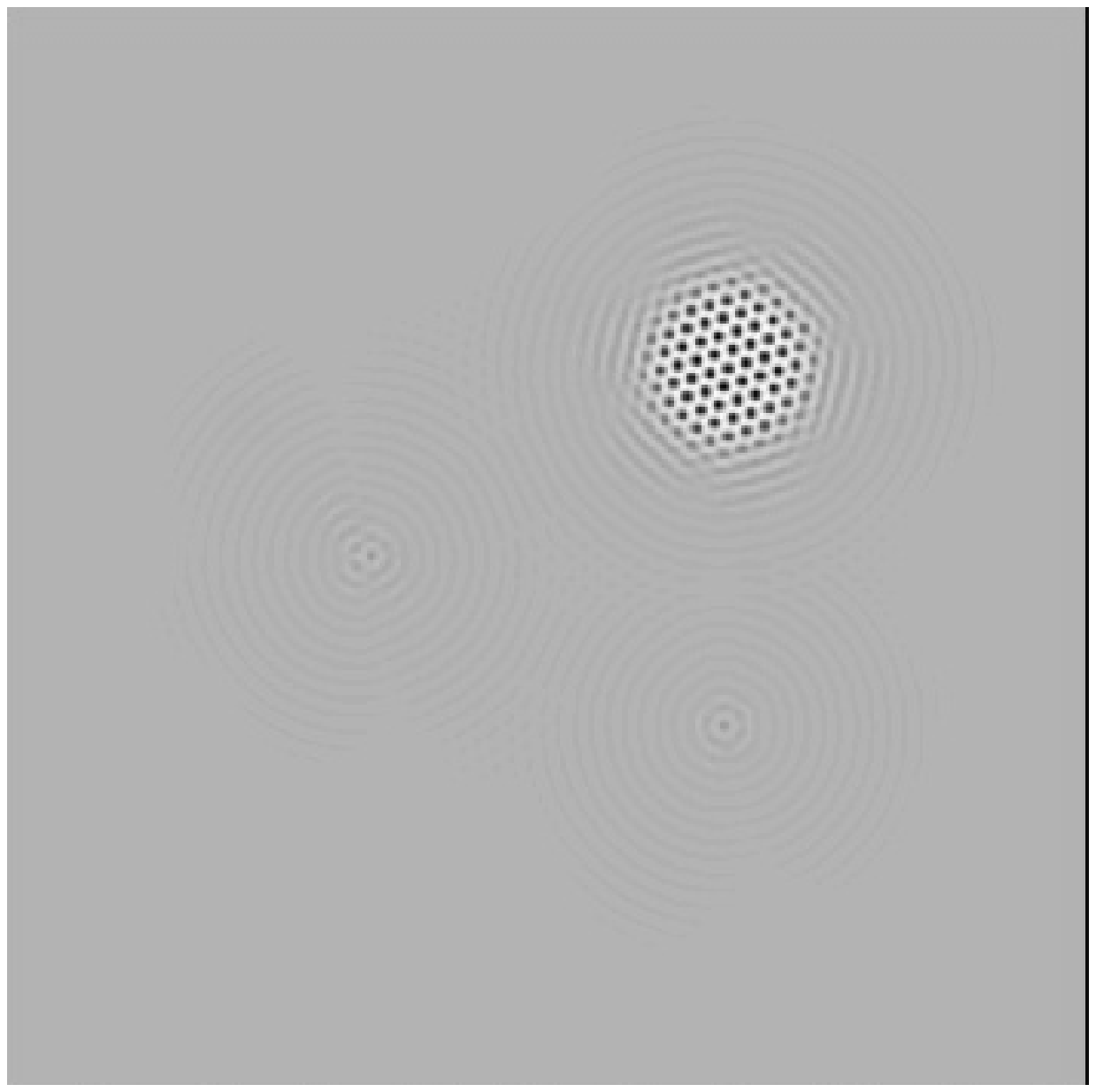}}
\subfigure[\label{fig:pfc2} $t=450$] {\includegraphics[height=2.5in,angle=0]{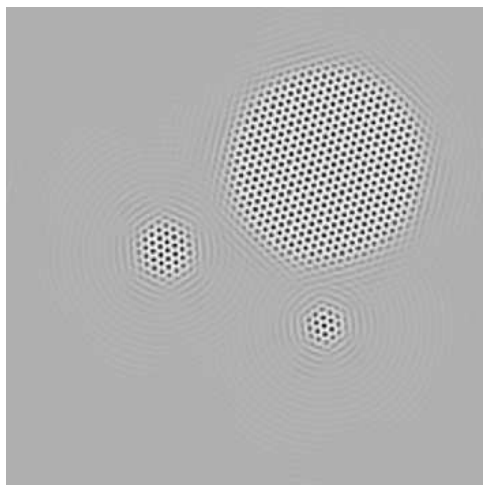}}
\subfigure[\label{fig:pfc3} $t=750$] {\includegraphics[height=2.5in,angle=0]{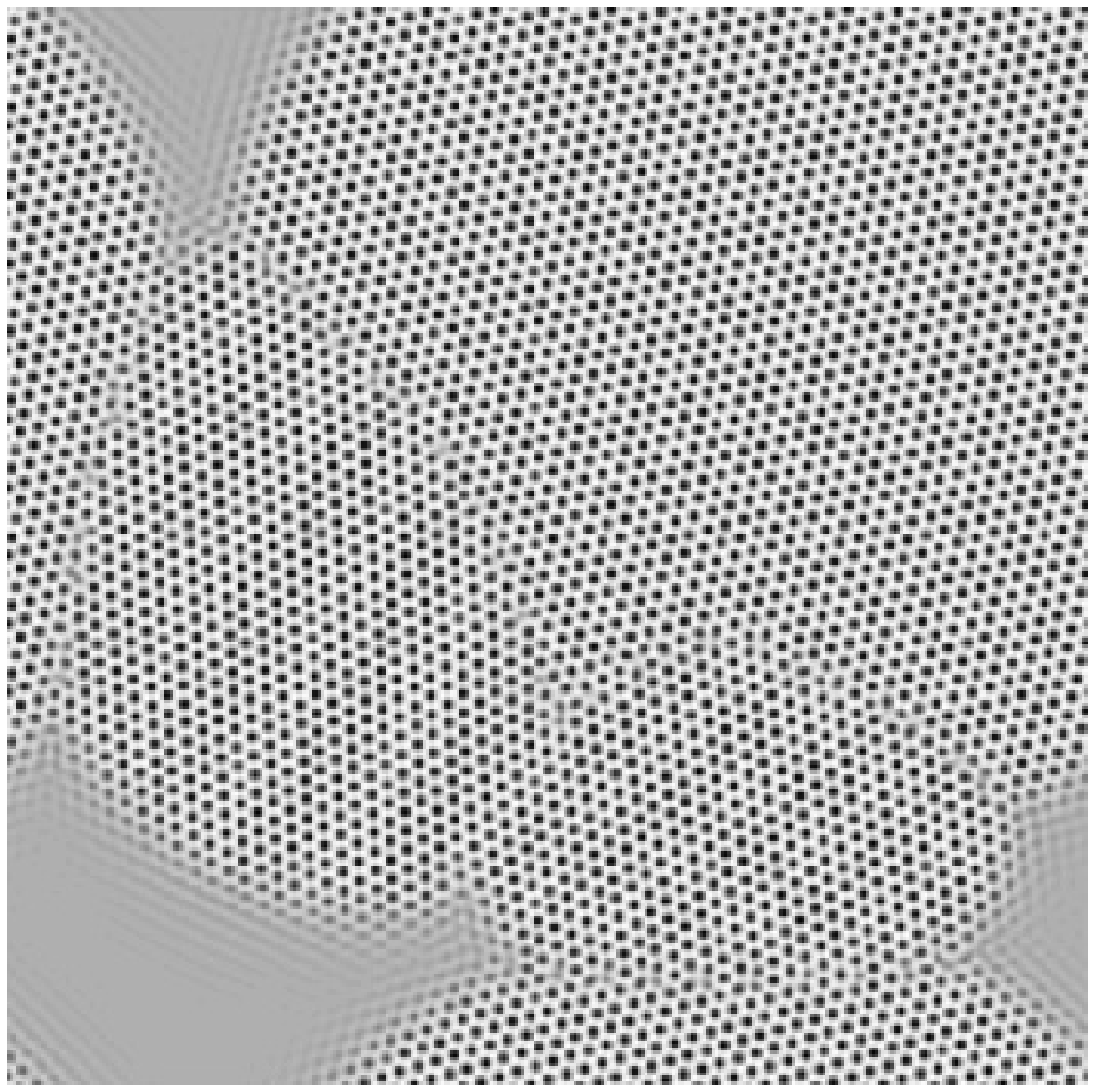}}
\caption{\label{fig:pfc} Heterogeneous nucleation, crystal growth, and formation of grain boundaries in a 2-D
film from three randomly oriented seeds, as simulated by the PFC model. The field plotted is the density
variable $\psi(\mathbf{x},t)$. Note that the pattern is periodic inside each grain.}
\end{center}
\end{figure}

For present purposes we will concern ourselves only with the uniform to
hexagonal phase transition (but are not in any way restricted to it),
and will disregard $\zeta$ in view of its relative unimportance in
describing phase transition kinetics. In a single mode approximation, a
hexagonal pattern is described by
\begin{equation}
\label{eqn:1mode} \psi(\mathbf{x},t)=\sum_{j=1}^3A_{j}(t)e^{i\mathbf{k}_j\cdot\mathbf{x}} + \bar{\psi} +
\mbox{c.c.},
\end{equation}
where $\mathbf{k}_1 = k_0 ({-\vec{\mbox{i}}\sqrt{3}/2} -
{\vec{\mbox{j}}/2})$, $\mathbf{k}_2 = k_0{{\vec{\mbox{j}}}}$ and
$\mathbf{k}_3 = k_0({\vec{\mbox{i}}\sqrt{3}/2} - {\vec{\mbox{j}}/2})$
are the reciprocal lattice vectors, $k_0$ is the wavenumber of the
pattern, $\vec{\mbox{i}}$ and $\vec{\mbox{j}}$ are unit vectors in the
$x$- and $y$-directions, $A_j$ are the complex amplitude functions, and
c.~c.~denotes complex conjugate. We point out that a striped pattern,
with the stripes parallel to the $x$ axis, can be represented by the
same equation above with $A_{1}=A_{3}=0$ and $A_{2}\ne0$. As shown in
Fig. \ref{fig:pfc} however, the PFC equation describes the evolution of
several such hexagonally patterned crystals of arbitrary orientation,
that collide to form grain boundaries. While the pattern remains
periodic within each crystal, there is a break in the periodicity
across the boundaries. Equation (\ref{eqn:1mode}) can be made to
describe such a system by allowing the $A_j$ to be spatially varying,
i.~e.~$A_{j}(\mathbf{x},t)$. Our goal is then to derive evolution
equations for $A_{j}(\mathbf{x},t)$, which along with Eq.
(\ref{eqn:1mode}) can be used to reconstruct $\psi(\mathbf{x},t)$ in a
one-mode approximation. Note that $A_{j}(\mathbf{x},t)$ now contains
information about both the envelope function (amplitude modulus), as
well as the orientation (phase angle) of each grain, but varies on a
much larger length scale.

In order to proceed with our analysis, which is essentially
perturbative, we identify $\epsilon$ as a small parameter, in powers of
which we shall expand $\psi$ about the steady state solution. We point
out that the PFC model stipulates no such restrictions on the value of
$\epsilon$, other than $\epsilon \ge 0$, whereas it is natural to
expect our treatment to restrict the validity of the amplitude
equations so derived to small values of $\epsilon \ll 1$.

\section{\label{QD}Heuristic RG calculation}
We now present a derivation of the amplitude equation from Eq.
(\ref{eqn:pfc}) using linear stability analysis and a shortcut
motivated by experience. An idea along these lines was previously
implemented by Bestehorn and Haken \cite{BH90} to derive an OPE
(similar to the Swift-Hohenberg equation) for modeling traveling waves
and pulses in two-dimensional systems, but not for deriving amplitude
equations.

We pose Eq. (\ref{eqn:pfc}) in a more convenient form by scaling $\psi$
by $\sqrt{\epsilon}$, and calling this new variable $\psi$. In this
manner, Eq. (\ref{eqn:pfc}) becomes
\begin{equation}
\label{eqn:pfc_scaled} \partial_t \psi =\nabla^2(1 + \nabla^2)^2\psi + \epsilon\nabla^2(\psi^3- \psi).
\end{equation}
Let us now consider the stability of the uniform phase solution $\bar{\psi}$ to the formation of the hexagonal
pattern by adding to it a small perturbation $\widetilde{\psi}$, so that $\psi = \bar{\psi} + \widetilde{\psi}$.
Substituting in Eq. (\ref{eqn:pfc_scaled}) and linearizing about $\bar{\psi}$ we obtain
\begin{equation}
\label{eqn:pfc_linearized} \partial_t \widetilde{\psi} =\nabla^2\left[\epsilon(3\bar{\psi}^2-1)+(1 +
\nabla^2)^2\right]\widetilde{\psi}
\end{equation}
If $\widetilde{\psi}$ is a hexagonal instability in the form given by
the spatially dependent part of Eq.~(\ref{eqn:1mode}), then using
$A_j(t)=A_{0j}\exp(\omega_j t)$, where $A_{0j}$ are complex constants,
and substituting in Eq. (\ref{eqn:pfc_linearized}), we obtain the
discrete dispersion relation
\begin{equation}
\label{eqn:ddr} \omega_{j} = -|\mathbf{k}_j|^2\left[\epsilon(3\bar{\psi}^2-1)+ \left(1-|\mathbf{k}_j|^2\right)^2
\right],
\end{equation}
after applying orthogonality conditions. Here $\omega_j$ predicts the
growth or decay rate of a hexagonal instability in the spatially
uniform system. Note that for real values of $\bar{\psi}$, $\omega_j$
is always real. Thus, a necessary condition for the instability to
grow, i.~e.~for $\omega_j$ to take on positive values, is
$3\bar{\psi}^2-1 < 0$, or equivalently $3\bar{\psi}^2-\epsilon < 0$ in
original variables. The most dangerous wave-number is the locus
$|\mathbf{k}_j| = k_0 = 1$.

We now consider spatial modulations in the amplitude about this preferred wave-number, i.~e.
\begin{equation} A_{j}(t) \longmapsto
A_{Rj}(\mathbf{x},t) = A_{0j}e^{\omega_{j}(\mathbf{Q})t}e^{i\mathbf{Q}\cdot\mathbf{x}},
\end{equation}
where $\mathbf{Q}=Q_x\;\vec{\mbox{i}}+Q_y\;\vec{\mbox{j}}$ is a
perturbation vector, and $A_{Rj}$ is the renormalized amplitude, whose
implication will be clarified in a following paragraph. Consistent with
Eq. (\ref{eqn:ddr}), we can now write the exponent controlling growth
rate along each lattice vector as
\begin{equation} \label{eqn:ddr_pert} \omega_{j}(\mathbf{Q}) =
|\mathbf{Q}+\mathbf{k}_j|^2\left[\epsilon(1-3\bar{\psi}^2) - \left(1-|\mathbf{Q}+\mathbf{k}_j|^2\right)^2
\right].
\end{equation}
We now replace the Fourier space variables in the above equation by their real space counterparts so that,
\begin{equation} \omega_j\equiv\partial_t\;\;,\;\;Q_x\equiv-i\partial_x\;\;,\;\;Q_y\equiv-i\partial_y\;\;,
\end{equation}
thus obtaining, \begin{equation} \label{eqn:ddr_oper}
|\mathbf{Q}+\mathbf{k}_j|^2\equiv1-\nabla^2-2i\mathbf{k}_j\cdot\nabla=1-\mathcal{L}_{\mathbf{k}_j}.
\end{equation}
Combining Eqs. (\ref{eqn:ddr_pert}) and (\ref{eqn:ddr_oper}), the space-time amplitude variations along each
lattice vector is given by the sixth order linear partial differential equation
\begin{eqnarray} \label{eqn:s-t_oper} \partial_t A_{Rj} +
(1-\mathcal{L}_{\mathbf{k}_j})\mathcal{L}_{\mathbf{k}_j}^2A_{Rj} &+&
\epsilon(1-3\bar{\psi}^2)\mathcal{L}_{\mathbf{k}_j}A_{Rj}\nonumber\\
&=& \epsilon(1-3\bar{\psi}^2)A_{Rj}.
\end{eqnarray}

We also need nonlinear terms, which play a vital role in pattern
dynamics near onset of the instability, to complement the above set of
equations. There are a couple of different ways to obtain these terms.
One can directly look for the nonlinear part in the normal form
equations \cite{Wigginsbook} for the dynamics of $A_j$ in a hexagonal
basis \cite{BuGol83,CCLPP90,gunaratne}. These equations have been
widely used to study the dynamics and stability of \emph{exactly}
periodic rolls and hexagonal patterns originating from the static
conducting state in Rayleigh-B\'{e}nard convection. Alternatively, one
can derive these terms to a particular order in $\epsilon$ through a
renormalization group (or multiple scales) analysis of the governing
differential equation. Here, we choose the latter approach, starting
from Eq. (\ref{eqn:pfc_scaled}), but only going far enough in the RG
analysis to identify the correct form of the terms.

We start with a perturbation series in $\epsilon$
\begin{equation}
\label{eqn:pseries} \psi = \psi_0 + \epsilon\psi_1 + \epsilon^2\psi_2 + \epsilon^3\psi_3 + \ldots
\end{equation}
where $\psi_0$ is a steady state solution and $\psi_{j(\ne0)}$ are the
higher order corrections. As we are interested in amplitude variations
in the hexagonal pattern, we pick $\psi_0$ to be the steady
\emph{hexagonal} solution, i.~e.~Eq.~(\ref{eqn:1mode}) with $A_j(t)$
replaced by $A_j(t\rightarrow\infty)$. Substituting in Eq.
(\ref{eqn:pfc_scaled}), we obtain the following equation at
$\mathcal{O}(\epsilon)$:
\begin{equation}
\label{eqn:PRG1} \left[\partial_t  - \nabla^2(1 + \nabla^2)^2\right]\psi_1 = \left(\partial_t -
\mathcal{L_P}\right)\psi_1 = \nabla^2(\psi_0^3- \psi_0),
\end{equation}
where
\begin{eqnarray}
\label{eqn:secterms} \nabla^2(\psi_0^3- \psi_0) & = &
(1-3\bar{\psi}^2)\sum_{j=1}^3A_{j}e^{i\mathbf{k}_j\cdot\mathbf{x}}\nonumber\\
& &\hspace{-2.5cm} -3A_{1}\left(|A_{1}|^2+2|A_{2}|^2+2|A_{3}|^2\right)e^{i\mathbf{k}_1\cdot\mathbf{x}}\nonumber\\
& &\hspace{-2.5cm} -3A_{2}\left(2|A_{1}|^2+|A_{2}|^2+2|A_{3}|^2\right)e^{i\mathbf{k}_2\cdot\mathbf{x}}\nonumber\\
& &\hspace{-2.5cm} -3A_{3}\left(2|A_{1}|^2+2|A_{2}|^2+|A_{3}|^2\right)e^{i\mathbf{k}_3\cdot\mathbf{x}}\nonumber\\
& &\hspace{-2.5cm} -6A_{2}^*A_{3}^*\bar{\psi}e^{i\mathbf{k}_1\cdot\mathbf{x}}
-6A_{1}^*A_{3}^*\bar{\psi}e^{i\mathbf{k}_2\cdot\mathbf{x}}\nonumber\\
& &\hspace{-2.5cm} -6A_{1}^*A_{2}^*\bar{\psi}e^{i\mathbf{k}_3\cdot\mathbf{x}} + \mbox{other terms} + \mbox{c.c.}
\end{eqnarray}
The superscript `*' denotes complex conjugation. To this order, the
``other terms'' are functions of complex exponentials that do not lie
in the null space of the linear differential operator in Eq.
(\ref{eqn:PRG1}), i.~e.~they are non-resonant terms. Therefore, they do
not contribute to unbounded growth in $\psi_1$. The terms listed in Eq.
(\ref{eqn:secterms}) are, however, resonant with the operator, and
their coefficients need to be renormalized in order to bound the
solution obtained by truncating the perturbation series at
$\mathcal{O}(\epsilon)$. The renormalization procedure allows the
amplitude $A_{j}$, previously constant, to now have space-time
variations that absorb secular divergences. We assert that the
nonlinear terms in the amplitude equation to $\mathcal{O}(\epsilon)$
must be the \emph{renormalized} coefficients of the exponential terms
in resonance with the differential operator. For example, the terms
complementing the space-time operator along basis vector $\mathbf{k}_1$
must be
\begin{eqnarray}
\label{eqn:nlnrterms} \epsilon(1-3\bar{\psi}^2)A_{R1}&-&3\epsilon
A_{R1}\left(|A_{R1}|^2+2|A_{R2}|^2+2|A_{R3}|^2\right)\nonumber\\
&-&6\epsilon A_{R2}^*A_{R3}^*\bar{\psi},
\end{eqnarray}
where the $A_{Rj}$ are the renormalized amplitude functions (no longer
constants). Note that these terms are completely identical to those
predicted by normal form theory for a hexagonal basis
\cite{gunaratne,CCLPP90}. Combining Eqs. (\ref{eqn:s-t_oper}) and
(\ref{eqn:nlnrterms}) we write the amplitude equation as
\begin{eqnarray}
\label{eqn:qdamp1} \partial_t A_1 &=& - (1-\mathcal{L}_{\mathbf{k}_1})\mathcal{L}_{\mathbf{k}_1}^2A_1 -
\epsilon(1-3\bar{\psi}^2)\mathcal{L}_{\mathbf{k}_1}A_1\nonumber\\
& &+\epsilon(1-3\bar{\psi}^2)A_1 - 3\epsilon A_1(|A_1|^2+2|A_2|^2+2|A_3|^2)\nonumber\\
& &-6\epsilon A_2^*A_3^*\bar{\psi},
\end{eqnarray}
for lattice vector $\mathbf{k}_1$, and permutations thereof for $\mathbf{k}_2$ and $\mathbf{k}_3$, where we have
replaced the variables $A_{Rj}$ by $A_j$.

We observe that the leading term in Eq. (\ref{eqn:nlnrterms}) is
consistent with the right hand side of Eq. (\ref{eqn:s-t_oper}),
thereby providing a natural overlapping link about which to match the
linear stability and perturbation results. In a more abstract sense, we
draw a parallel between this method and the technique of matched
asymptotic expansions in singular perturbation theory, where inner and
outer asymptotic solutions are matched over a common region of validity
in the solution space, to obtain a globally valid solution. This
completes our derivation of the amplitude equation via a heuristic or
``quick and dirty'' approach. For future reference, we will call Eq.
(\ref{eqn:qdamp1}), the QDRG (quick and dirty RG) equation, and the
method used to obtain it as the QDRG (or heuristic) method. As we have
already demonstrated the remarkable accuracy with which the QDRG
equation mimics the PFC equation in a previous article \cite{GAD05_1},
we will refrain from presenting any new evidence to that effect here.

To summarize the procedure, we first conducted a linear stability
analysis of the scaled PFC equation about the uniform state to obtain a
linear differential operator controlling the space-time evolution of
the complex amplitude $A_j$ of the hexagonal pattern. We superimposed
on this dispersion relation periodic modulations of the amplitude, and
from the dispersion relation in terms of these latter modulations,
identified the gradient terms in the amplitude equation. We then
carried out the first step in a conventional RG analysis to obtain the
form of the nonlinear terms that should accompany this differential
operator, and combining the two results, we wrote down the amplitude
equation for the hexagonal pattern. In this respect, our approach lacks
the full mathematical rigor of a conventional RG reduction or a
multiple scales derivation, which gives it a somewhat ``dirty''
appearance. However, we made no assumptions about the scaling of the
space-time variables in the system, nor did we have to construct any
secular solutions so far. We will comment on extending this method
systematically to higher orders in $\epsilon$ in the following section.

\section{\label{PRG}Proto-Renormalization group derivation}
With the proto-RG method, our starting point is Eq.~(\ref{eqn:pfc_scaled}) with the perturbation series
Eq.~(\ref{eqn:pseries}). Thus, to $\mathcal{O}(\epsilon)$ we obtain Eq.~(\ref{eqn:PRG1}), whereas to
$\mathcal{O}(\epsilon^2)$ we get
\begin{equation}
\label{eqn:PRG2} \left(\partial_t-\mathcal{L_P}\right)\psi_2 = \nabla^2(3\psi_0^2\psi_1- \psi_1).
\end{equation}
The structure of Eq.~(\ref{eqn:PRG1}) allows us to infer that its simplest particular solution will take the
form
\begin{eqnarray}
\label{eqn:PRG1_PS} \psi_1 &=& \sum_{j=1}^3P_{1j}(\mathbf{x},t)e^{i\mathbf{k}_j\cdot\mathbf{x}}
+ \sum_{j=1}^3Q_{1j}e^{2i\mathbf{k}_j\cdot\mathbf{x}}\nonumber\\
& & + \sum_{j=1}^3R_{1j}e^{3i\mathbf{k}_j\cdot\mathbf{x}} +
\sum_{j=1}^3S_{1j}e^{i\mathbf{s}_j\cdot\mathbf{x}}\nonumber\\
& & + \sum_{j=1}^2T_{1j}e^{i\mathbf{t}_j\cdot\mathbf{x}} +
\sum_{j=1}^2U_{1j}e^{i\mathbf{u}_j\cdot\mathbf{x}}\nonumber\\
& & + \sum_{j=1}^2V_{1j}e^{i\mathbf{v}_j\cdot\mathbf{x}} + \mbox{c. c.},
\end{eqnarray}
where
\begin{eqnarray*}
\mathbf{s}_1 = -\vec{\mbox{i}}\;\frac{\sqrt{3}}{2} - \vec{\mbox{j}}\;\frac{3}{2}, \mathbf{s}_2 =
\vec{\mbox{i}}\;\frac{\sqrt{3}}{2} - \vec{\mbox{j}}\;\frac{3}{2}, \mathbf{s}_3 = \mathbf{s}_2-\mathbf{s}_1
\end{eqnarray*}
\begin{eqnarray*}
\mathbf{t}_1=-\vec{\mbox{i}}\;\frac{3\sqrt{3}}{2} -
\vec{\mbox{j}}\;\frac{1}{2},\mathbf{t}_2=\vec{\mbox{i}}\;\frac{3\sqrt{3}}{2} - \vec{\mbox{j}}\;\frac{1}{2}
\end{eqnarray*}
\begin{eqnarray*}
\mathbf{u}_1=-\vec{\mbox{i}}\;\frac{\sqrt{3}}{2} -
\vec{\mbox{j}}\;\frac{5}{2},\mathbf{u}_2=\vec{\mbox{i}}\;\frac{\sqrt{3}}{2} - \vec{\mbox{j}}\;\frac{5}{2}
\end{eqnarray*}
\begin{eqnarray}
\label{eqn:PRG1_vectors} \mathbf{v}_1=-\vec{\mbox{i}}\;\sqrt{3} -
\vec{\mbox{j}}\;2,\mathbf{v}_2=\vec{\mbox{i}}\;\sqrt{3} - \vec{\mbox{j}}\;2,
\end{eqnarray}
are non-resonant modes generated by the cubic term. Note that we have
explicitly denoted the space-time dependence of the secular
coefficients $P_{1j}(\mathbf{x},t)$, which are polynomials in $x$, $y$
and $t$, whereas by inspection, the other coefficients $Q_{1j}$,
$R_{1j}$, $S_{1j}$, $T_{1j}$, $U_{1j}$ and $V_{1j}$ \emph{can} be
complex constants. Specifically, $P_{11}$ satisfies
\begin{eqnarray*}
\left(\partial_t-\mathcal{L_P}\right)P_{11}e^{i\mathbf{k}_1\cdot\mathbf{x}}&=&(1-3\bar{\psi}^2)A_{1}e^{i\mathbf{k}_1\cdot\mathbf{x}}\nonumber\\
& &\hspace{-2.5cm} -3A_{1}\left(|A_{1}|^2+2|A_{2}|^2+2|A_{3}|^2\right)e^{i\mathbf{k}_1\cdot\mathbf{x}}\nonumber\\
& &\hspace{-2.5cm} -6A_{2}^*A_{3}^*\bar{\psi}e^{i\mathbf{k}_1\cdot\mathbf{x}}
\end{eqnarray*}
\begin{eqnarray}
\label{eqn:part_PRG_eps} \Rightarrow \left(\partial_t +
(1-\mathcal{L}_{\mathbf{k}_1})\mathcal{L}_{\mathbf{k}_1}^2\right)P_{11}
&=&(1-3\bar{\psi}^2)A_{1}\nonumber\\
& &\hspace{-4.5cm} -3A_{1}\left(|A_{1}|^2+2|A_{2}|^2+2|A_{3}|^2\right)
-6A_{2}^*A_{3}^*\bar{\psi}\nonumber\\
&\equiv& L_{\mathbf{k}_1}P_{11},
\end{eqnarray}
where $L_{\mathbf{k}_j}$ is the ``proto-RG'' operator for lattice vector $\mathbf{k}_j$. From the above equation
it is quite obvious that $P_{11}$ cannot be constant for any non-trivial solutions, and likewise for $P_{12}$
and $P_{13}$.

As $P_{1j}$ are secular, we now renormalize \cite{CGO2} $\psi$ about arbitrary regularization points
$\mathbf{X}$ and $T$, as in the conventional RG method, to get
\begin{eqnarray}
\label{eqn:renorm_pseries} \psi &=& \bar{\psi}+
\sum_{j=1}^3A_{Rj}(\mathbf{X},T)e^{i\mathbf{k}_j\cdot\mathbf{x}}\nonumber\\
& &+
\epsilon\sum_{j=1}^3\left(P_{1j}(\mathbf{x},t)-P_{1j}(\mathbf{X},T)\right)e^{i\mathbf{k}_j\cdot\mathbf{x}}\nonumber\\
& &+ \ldots + \mbox{c.c.}
\end{eqnarray}
where $A_{Rj}$ is now the renormalized amplitude that absorbs secular divergences. Since $\psi$ must be
independent of these regularization points, we have
\begin{eqnarray}
\label{eqn:PRG_eps}
L_{\mathbf{k}_1}^{\mathbf{X},T}\psi &=& 0\nonumber\\
\Rightarrow L_{\mathbf{k}_1}^{\mathbf{X},T}A_{R1}(\mathbf{X},T) &=& \epsilon
L_{\mathbf{k}_1}^{\mathbf{X},T}P_{11}(\mathbf{X},T)
+ \epsilon^2 L_{\mathbf{k}_1}^{\mathbf{X},T}P_{21}(\mathbf{X},T)\nonumber\\
& & + \epsilon^3 L_{\mathbf{k}_1}^{\mathbf{X},T}P_{31}(\mathbf{X},T) + \ldots
\end{eqnarray}
after applying orthogonality conditions. This is the general form of
the proto-RG equation for weakly nonlinear oscillators \cite{Nozaki01}.
$L_{\mathbf{k}_1}^{\mathbf{X},T}$ is the proto-RG operator
$L_{\mathbf{k}_1}$ in Eq.~(\ref{eqn:part_PRG_eps}), with variables
$\mathbf{x}$ and $t$ replaced by $\mathbf{X}$ and $T$ respectively.
Changing back from $(\mathbf{X},T)\to(\mathbf{x},t)$ and $A_{Rj}\to
A_j$, and using Eqs.~(\ref{eqn:part_PRG_eps}) and (\ref{eqn:PRG_eps}),
we can write the amplitude equation along lattice vector $\mathbf{k}_1$
to $\mathcal{O}(\epsilon)$ explicitly as
\begin{eqnarray}
\label{eqn:full_PRG_eps} \partial_t A_1 &=& -(1-\mathcal{L}_{\mathbf{k}_1})\mathcal{L}_{\mathbf{k}_1}^2A_1 +
\epsilon(1-3\bar{\psi}^2)A_1-3\epsilon
A_1(|A_1|^2\nonumber\\
& &+2|A_2|^2+2|A_3|^2)-6\epsilon A_2^*A_3^*\bar{\psi},
\end{eqnarray}
with appropriate permutations for $A_2$ and $A_3$. Note that in using
Eq.~(\ref{eqn:part_PRG_eps}), we have replaced $A_{j}$ by their
renormalized counterparts $A_{Rj}$ as is consistent with the proto-RG
procedure, before reverting to the former notation for amplitude. Upon
comparing the two amplitude equations obtained so far,
Eqs.~(\ref{eqn:qdamp1}) and (\ref{eqn:full_PRG_eps}), we note that the
QDRG derived equation carries the extra term
$\epsilon(1-3\bar{\psi}^2)\mathcal{L}_{\mathbf{k}_1}A_1$. Evidently,
the QDRG and the proto-RG methods produce different amplitude equations
when applied to the PFC equation, the extent of this difference
controlled by the parameter $\epsilon(1-3\bar{\psi}^2)$.

As mentioned earlier, the principal advantage of using the proto-RG
method is the relative ease with which one can progress to higher order
calculations. Let us now extend this calculation to
$\mathcal{O}(\epsilon^2)$. We need $\psi_1$ in order to evaluate the
right hand side of Eq.~(\ref{eqn:PRG2}), which means that we
additionally need to evaluate $Q_{1j}$, $R_{1j}$, $S_{1j}$, $T_{1j}$,
$U_{1j}$ and $V_{1j}$. The constant values of these terms can be
determined by inspection. For example, by analogy with $P_{11}$, we see
that $Q_{11}$ must satisfy
\begin{eqnarray*}
\left(\partial_t-\mathcal{L_P}\right)Q_{11}e^{2i\mathbf{k}_1\cdot\mathbf{x}}&=&
-12\left(A_1^2\bar{\psi}+2A_1A_2^*A_3^*\right)e^{2i\mathbf{k}_1\cdot\mathbf{x}}
\end{eqnarray*}
\begin{eqnarray}
\label{eqn:q1solve} \Rightarrow \left(\partial_t +
(4-\mathcal{L}_{2\mathbf{k}_1})(\mathcal{L}_{2\mathbf{k}_1}-3)^2\right)Q_{11}
&=&\nonumber\\
& &\hspace{-3.5cm}-12\left(A_1^2\bar{\psi}+2A_1A_2^*A_3^*\right).
\end{eqnarray}
Unlike Eq.~(\ref{eqn:part_PRG_eps}) however, we see that
Eq.~(\ref{eqn:q1solve}) permits a constant solution for $Q_{11}$, which
in turn is determined to be
\begin{equation}
\label{eqn:q1soln} Q_{11} = -\frac{1}{3}\left(A_1^2\bar{\psi}+2A_1A_2^*A_3^*\right).
\end{equation}
Similarly, constant solutions for the other coefficients are
\begin{eqnarray}
\label{eqn:allsolns} Q_{12} &=&
-\frac{1}{3}\left(A_2^2\bar{\psi}+2A_1^*A_2A_3^*\right)\nonumber\\
Q_{13} &=&
-\frac{1}{3}\left(A_3^2\bar{\psi}+2A_1^*A_2^*A_3\right)\nonumber\\
R_{1j} &=& -\frac{A_j^3}{64}\nonumber\\
S_{11} &=&
-\frac{3}{4}\left(A_1^2A_3+2A_1\bar{\psi}A_2^*+{A_2^{*}}^2A_3^*\right)\nonumber\\
S_{12} &=&
-\frac{3}{4}\left(A_1A_3^2+2A_2^*\bar{\psi}A_3+{A_2^{*}}^2A_1^*\right)\nonumber\\
S_{13} &=&
-\frac{3}{4}\left(A_2A_3^2+2A_1^*\bar{\psi}A_3+{A_1^{*}}^2A_2^*\right)\nonumber\\
T_{11} &=& -\frac{A_1^2A_3^*}{12}\nonumber\\
T_{12} &=& -\frac{A_3^2A_1^*}{12}\nonumber\\
U_{11} &=& -\frac{A_1{A_2^*}^2}{12}\nonumber\\
U_{12} &=& -\frac{A_3{A_2^*}^2}{12}\nonumber\\
V_{11} &=& -\frac{A_1^2A_2^*}{12}\nonumber\\
V_{12} &=& -\frac{A_3^2A_2^*}{12}.
\end{eqnarray}
We know that the particular solution to Eq.~(\ref{eqn:PRG2}) has the form
\begin{equation}
\psi_2 = \sum_{j=1}^3P_{2j}(\mathbf{x},t)e^{i\mathbf{k}_j\cdot\mathbf{x}} + \ldots + \mbox{c. c.}
\end{equation}
where we have shown only the resonant part of the solution. The terms on the right hand side of
Eq.~(\ref{eqn:PRG2}), resonant with lattice vector $\mathbf{k}_1$, evaluate to
\begin{eqnarray}
\Phi&=&[\left(1-3\bar{\psi}^2-6(|A_1|^2+|A_2|^2+|A_3|^2)\right)\left(1-\mathcal{L}_{\mathbf{k}_1}\right)P_{11}\nonumber\\
& &-3A_1^2\left(1-\mathcal{L}_{\mathbf{k}_1}\right)P_{11}^*-6A_1A_2\left(1-\mathcal{L}_{\mathbf{k}_1}\right)P_{12}^*\nonumber\\
& &-6A_1A_3\left(1-\mathcal{L}_{\mathbf{k}_1}\right)P_{13}^*-6\bar{\psi}A_3^*\left(1-\mathcal{L}_{\mathbf{k}_1}\right)P_{12}^*\nonumber\\
& &-6\bar{\psi}A_2^*\left(1-\mathcal{L}_{\mathbf{k}_1}\right)P_{13}^*-6A_1A_2^*\left(1-\mathcal{L}_{\mathbf{k}_1}\right)P_{12}\nonumber\\
& &-6A_1A_3^*\left(1-\mathcal{L}_{\mathbf{k}_1}\right)P_{13}-6A_2\bar{\psi}S_{11}-3A_2^2U_{11}\nonumber\\
& &-6A_3A_1^*T_{11}-6A_2A_1^*V_{11}-6A_1^*A_3^*S_{11}-3{A_3^*}^2S_{12}\nonumber\\
& &-3{A_2^*}^2S_{12}^*-6A_3\bar{\psi}S_{13}^*-6A_1^*A_2^*S_{13}^*-3A_3^2T_{12}^*\nonumber\\
& &-6A_2A_3^*Q_{12}^*-6A_3A_2^*Q_{13}^*-3{A_1^*}^2R_{11}\nonumber\\
& &-6A_2A_3Q_{11}-6\bar{\psi}A_1^*Q_{11}]e^{i\mathbf{k}_1\cdot\mathbf{x}}.
\end{eqnarray}
Thus $P_{21}$ satisfies
\begin{equation}
\label{eqn:part_PRG_eps2} L_{\mathbf{k}_1}P_{21} = \Phi.
\end{equation}
The non-constant terms in $\Phi$ (terms containing $P_{1j}$) are now
ignored \cite{Nozaki00,Nozaki01} while the remaining terms are
determined from their constant solutions, Eqs. (\ref{eqn:q1soln}) and
(\ref{eqn:allsolns}). Thus, using Eq.~(\ref{eqn:PRG_eps}) we can write
the amplitude equation along lattice vector $\mathbf{k}_1$ to
$\mathcal{O}(\epsilon^2)$ as
\begin{eqnarray}
\label{eqn:full_PRG_eps2} \partial_t A_1 &=& -(1-\mathcal{L}_{\mathbf{k}_1})\mathcal{L}_{\mathbf{k}_1}^2A_1 +
\epsilon(1-3\bar{\psi}^2)A_1-3\epsilon
A_1(|A_1|^2\nonumber\\
& &+2|A_2|^2+2|A_3|^2)-6\epsilon
A_2^*A_3^*\bar{\psi}+11\epsilon^2\bar{\psi}A_1^2A_2A_3\nonumber\\
& &+\epsilon^2\bar{\psi}^2A_1(2|A_1|^2+9|A_2|^2+9|A_3|^2)\nonumber\\
& &+11\epsilon^2\bar{\psi}(2|A_1|^2+|A_2|^2+|A_3|^2)A_2^*A_3^*\nonumber\\
& &+\frac{27}{2}\epsilon^2A_1^*{A_2^*}^2{A_3^*}^2+5\epsilon^2A_1|A_1|^2\left(|A_2|^2+|A_3|^2\right)\nonumber\\
&
&+12\epsilon^2A_1|A_2|^2|A_3|^2+\frac{3}{64}\epsilon^2A_1|A_1|^4\nonumber\\
& &+\frac{5}{2}\epsilon^2A_1|A_2|^4+\frac{5}{2}\epsilon^2A_1|A_3|^4,
\end{eqnarray}
with cyclic permutations for lattice vectors $\mathbf{k}_2$ and $\mathbf{k}_3$.

We can in principle extend the QDRG method also to higher orders by
performing the same steps above, until the point where we identify the
resonant terms on the right hand side of Eq.~(\ref{eqn:PRG2}),
i.e.~$\Phi$. Combining this result with Eqs.~(\ref{eqn:s-t_oper}) and
(\ref{eqn:nlnrterms}) we can then obtain the amplitude equation
Eq.(\ref{eqn:full_PRG_eps2}), but with an extra term
$\epsilon(1-3\bar{\psi}^2)\mathcal{L}_{\mathbf{k}_1}A_1$.

In summary, both the proto-RG and the QDRG can be calculated including
terms of $O(\epsilon^2)$, and the results differ by a small but
non-zero term.  Which, if any, of these calculations is correct?  And
what is the origin of the discrepancy between the two methods?  Is the
QDRG result not to be trusted, being derived heuristically? Faced with
two seemingly incompatible, although very similar results, it is
natural to attempt an independent test of the analysis, which we did
using the standard method of multiple scales.  This calculation is
presented below, but owing to technical complications arising from the
interference of modes and the need to go to sixth order of perturbation
theory, we found it only feasible to perform the calculation for the
case of one dimension.  Nevertheless, we will see that, in fact, the
QDRG result, Eq. (\ref{eqn:qdamp1}), is \emph{more} correct. The small
discrepancy between this result and the proto-RG result is finally
resolved in Section \ref{MODI}.

\section{\label{MS}Multiple scales derivation}
We now re-derive the amplitude equation using the traditional method of
multiple scales. As the primary purpose of this derivation is to verify
the previous calculations via an independent method, we stick to a one
dimensional analysis here that considerably simplifies the algebra. For
convenience we use $\delta^2 = \epsilon$, and write
Eq.~(\ref{eqn:pfc_scaled}) in 1-D as
\begin{equation}
\label{eqn:pfc_scaled_1d} \left[\partial_t -
\partial_x^2\left(1 +
\partial_x^2\right)^2\right]\psi =
\delta^2\partial_x^2(\psi^3- \psi).
\end{equation}
The basic premise of the multiple scales analysis is that while the
pattern itself varies on the scale of its wavelength ($2\pi/k_0$), its
amplitude varies on much larger length and time scales. It is then
appropriate to introduce \emph{slowly} varying arguments
\begin{equation}
\label{eqn:msxt}
 X = \delta x\;\;\;,\;\;\;T = \delta^2 t
\end{equation}
for the envelope function $A(X,T)$. This scaling was previously applied
by Gunaratne \emph{et al.} \cite{gunaratne} to the Swift-Hohenberg
equation with success (based on the form of the discrete dispersion
relation), and as the PFC equation is essentially a conserved analog of
the Swift-Hohenberg equation we anticipate that the same scaling holds
here.

Derivatives scale as follows
\begin{eqnarray}
\label{eqn:msdxdt} \partial_x &\to& \partial_x +
\delta\partial_X\nonumber\\
\partial_x^2 &\to& \partial_x^2 +
2\delta\partial_X\partial_x +
 \delta^2\partial_X^2\nonumber\\
\partial_t &\to& \delta^2\partial_T,
\end{eqnarray}
whereas the operator
\begin{eqnarray}
\label{eqn:msoper1} \partial_x^2\left(1 +
\partial_x^2\right)^2 &\to& \sum_{j=0}^6\delta^j\mathcal{L}_j
\end{eqnarray}
such that
\begin{eqnarray}
\label{eqn:msoper2} \mathcal{L}_0 &=& \partial_x^2\left(1 +
\partial_x^2\right)^2\nonumber\\
\mathcal{L}_1 &=& 4\partial_X\partial_x^3\left(1 +
\partial_x^2\right)+2\partial_X\partial_x\left(1 +
\partial_x^2\right)^2\nonumber\\
\mathcal{L}_2 &=& 4\partial_X^2\partial_x^4+10\partial_X^2\partial_x^2\left(1
+\partial_x^2\right)+\partial_{X}^2\left(1 +
\partial_x^2\right)^2\nonumber\\
\mathcal{L}_3 &=& 12\partial_X^3\partial_x^3+8\partial_X^3\partial_x\left(1
+\partial_x^2\right)\nonumber\\
\mathcal{L}_4 &=& 13\partial_X^4\partial_x^2+2\partial_X^4\left(1
+\partial_x^2\right)\nonumber\\
\mathcal{L}_5 &=& 6\partial_X^5\partial_x\nonumber\\
\mathcal{L}_6 &=& \partial_X^6.
\end{eqnarray}
We now expand $\psi$ in a perturbation series in $\delta$ to get
\begin{equation} \label{eqn:mspseries} \psi = \psi_0 +
\delta\psi_1 + \delta^2\psi_2 + \delta^3\psi_3 + \ldots.
\end{equation}
Using Eq.~(\ref{eqn:msdxdt}) and the above series, the $\delta$ expansion of the nonlinear term in
Eq.~(\ref{eqn:pfc_scaled_1d}) can be written as
\begin{eqnarray}
\label{eqn:msnlnr}
\partial_x^2(\psi^3-\psi) &=& \partial_x^2(\psi_0^3-\psi_0)\nonumber\\
&
&+\delta[\partial_x^2(3\psi_0^2\psi_1-\psi_1)+2\partial_X\partial_x(\psi_0^3-\psi_0)]\nonumber\\
&
&+\delta^2[\partial_x^2(3\psi_0\psi_1^2+3\psi_0^2\psi_2-\psi_2)\nonumber\\
&&+2\partial_X\partial_x(3\psi_0^2\psi_1-\psi_1)+\partial_X^2(\psi_0^3-\psi_0)]\nonumber\\
&
&+\delta^3[\partial_x^2(\psi_1^2+6\psi_0\psi_1\psi_2+3\psi_0^2\psi_3-\psi_3)\nonumber\\
&
&+2\partial_X\partial_x(3\psi_0\psi_1^2+3\psi_0^2\psi_2-\psi_2)\nonumber\\
& &+\partial_X^2(3\psi_0^2\psi_1-\psi_1)]+\delta^4[\partial_x^2(3\psi_1^2\psi_2\nonumber\\
& &+3\psi_0\psi_2^2+6\psi_0\psi_1\psi_3+3\psi_0^2\psi_4-\psi_4)\nonumber\\
& &+2\partial_X\partial_x(\psi_1^2+6\psi_0\psi_1\psi_2+3\psi_0^2\psi_3-\psi_3)\nonumber\\
&
&+\partial_X^2(3\psi_0\psi_1^2+3\psi_0^2\psi_2-\psi_2)]\nonumber\\
& &+ \mathcal{O}(\delta^5).
\end{eqnarray} Substituting Eq.~(\ref{eqn:mspseries}) in
Eq.~(\ref{eqn:pfc_scaled_1d}), and using the scaled operators in
Eqs.~(\ref{eqn:msdxdt}-\ref{eqn:msoper2}), we can write equations
satisfied by the $\psi_m$ at each $\mathcal{O}(\delta^m)$. At
$\mathcal{O}(1)$ we obtain,
\begin{eqnarray}
\label{eqn:ms1} \mathcal{L}_0\psi_0 &=& 0\nonumber\\
\Rightarrow \psi_0 &=& \bar{\psi} + A_{01}(X,T)e^{ix} + \mbox{c.c.}
\end{eqnarray}
where $A_{mn}$ is the complex amplitude of mode $n$ at
$\mathcal{O}(\delta^m)$. At $\mathcal{O}(\delta)$ we get
\begin{eqnarray}
\label{eqn:ms2}
\mathcal{L}_0\psi_1 + \mathcal{L}_1\psi_0 &=& 0\nonumber\\
\Rightarrow \psi_1 &=& A_{11}(X,T)e^{ix} + \mbox{c.c.}
\end{eqnarray}
where (and henceforth) we neglect the constant term in view of its
inclusion in Eq.~(\ref{eqn:ms1}). At the next order we have
\begin{equation}
\label{eqn:ms3} \mathcal{L}_0\psi_2 =
\partial_T\psi_0-\mathcal{L}_1\psi_1-\mathcal{L}_2\psi_0-\partial_{x}^2(\psi_0^3-\psi_0).
\end{equation}
For $\psi_2(x,t)$ to remain bounded we have to guarantee that the right
hand side of Eq.~(\ref{eqn:ms3}) does not have a projection in the null
space of $\mathcal{L}_0$, which yields a solvability condition
\cite{Orszagbook,Nayfehbook} (also known as the Fredholm alternative).
Applying the alternative imposes the following condition on the
amplitude at $\mathcal{O}(\delta^2)$:
\begin{equation}
\label{eqn:msamp3}
\partial_TA_{01} = 4\partial_X^2A_{01} +
(1-3\bar{\psi}^2)A_{01}-3A_{01}|A_{01}|^2.
\end{equation}
Thus,
\begin{equation}
\label{eqn:mspsi2} \psi_2 = A_{21}e^{ix}+A_{22}e^{2ix}+A_{23}e^{3ix}+\mbox{c.c.}
\end{equation}
where $A_{22} = A_{01}^2\bar{\psi}/3$, and $A_{23} = A_{01}^3/64$.

At subsequent orders, the following equations are obtained for $\psi_m$:
\begin{eqnarray}
\label{eqn:ms4to7} \mathcal{O}(\delta^3): \mathcal{L}_0\psi_3 &=&
\partial_T\psi_1-\mathcal{L}_1\psi_2-\mathcal{L}_2\psi_1-\mathcal{L}_3\psi_0\nonumber\\
&
&-[\partial_x^2(3\psi_0^2\psi_1-\psi_1)+2\partial_X\partial_x(\psi_0^3-\psi_0)]\nonumber\\
\mathcal{O}(\delta^4): \mathcal{L}_0\psi_4 &=&
\partial_T\psi_2-\mathcal{L}_1\psi_3-
\mathcal{L}_2\psi_2-\mathcal{L}_3\psi_1-\mathcal{L}_4\psi_0\nonumber\\
& &-[\partial_x^2(3\psi_0\psi_1^2+3\psi_0^2\psi_2-\psi_2)\nonumber\\
&&+2\partial_X\partial_x(3\psi_0^2\psi_1-\psi_1)+\partial_X^2(\psi_0^3-\psi_0)]\nonumber\\
\mathcal{O}(\delta^5): \mathcal{L}_0\psi_5 &=&
\partial_T\psi_3-\mathcal{L}_1\psi_4-
\mathcal{L}_2\psi_3-\mathcal{L}_3\psi_2-\mathcal{L}_4\psi_1\nonumber\\
&&-\mathcal{L}_5\psi_0-[\partial_x^2(\psi_1^2+6\psi_0\psi_1\psi_2+3\psi_0^2\psi_3\nonumber\\
&
&-\psi_3)+2\partial_X\partial_x(3\psi_0\psi_1^2+3\psi_0^2\psi_2-\psi_2)\nonumber\\
& &+\partial_X^2(3\psi_0^2\psi_1-\psi_1)]\nonumber\\
\mathcal{O}(\delta^6): \mathcal{L}_0\psi_6 &=&
\partial_T\psi_4-\mathcal{L}_1\psi_5-
\mathcal{L}_2\psi_4-\mathcal{L}_3\psi_3-\mathcal{L}_4\psi_2\nonumber\\
&&-\mathcal{L}_5\psi_1-\mathcal{L}_6\psi_0-[\partial_x^2(3\psi_1^2\psi_2+3\psi_0\psi_2^2\nonumber\\
& &+6\psi_0\psi_1\psi_3+3\psi_0^2\psi_4-\psi_4)+2\partial_X\partial_x(\psi_1^2\nonumber\\
&
&+6\psi_0\psi_1\psi_2+3\psi_0^2\psi_3-\psi_3)+\partial_X^2(3\psi_0\psi_1^2\nonumber\\
& &+3\psi_0^2\psi_2-\psi_2)],
\end{eqnarray}
and successive applications of the Fredholm alternative yield the following amplitude equations at those
respective orders,
\begin{eqnarray}
\label{eqn:msamp4to7}
\partial_TA_{11} &=& - 12i\partial_X^3A_{01} + 4\partial_X^2A_{11} - (1-3\bar{\psi}^2)2i\partial_XA_{01}\nonumber\\
&
&+(1-3\bar{\psi}^2)A_{11}-6A_{11}|A_{01}|^2-3A_{01}^2A_{11}^*\nonumber\\
& &+6i\partial_X(A_{01}^2A_{01}^*)\nonumber\\
\partial_TA_{21} &=& -13\partial_X^4A_{01}-12i\partial_X^3A_{11} +
4\partial_X^2A_{21}\nonumber\\
&
&-(1-3\bar{\psi}^2)(2i\partial_XA_{11}+\partial_X^2A_{01})\nonumber\\
&
&+(1-3\bar{\psi}^2)A_{21}-3A_{11}^2A_{01}^*-6|A_{01}|^2A_{21}\nonumber\\
& &-6\bar{\psi}A_{22}A_{01}^*-3A_{23}{A_{01}^*}^2-6A_{01}|A_{11}|^2\nonumber\\
&
&-3A_{01}^2A_{21}^*+6i\partial_X(A_{01}^2A_{11}^*+2|A_{01}|^2A_{11})\nonumber\\
& &+3\partial_X^2(A_{01}^2A_{01}^*)\nonumber\\
\partial_TA_{31} &=& 6i\partial_X^5A_{01}-13\partial_X^4A_{11}-12i\partial_X^3A_{21} +
4\partial_X^2A_{31}\nonumber\\
& &-(1-3\bar{\psi}^2)(2i\partial_XA_{21}+\partial_X^2A_{11})\nonumber\\
&
&+(1-3\bar{\psi}^2)A_{31}-6A_{11}A_{21}A_{01}^*-6|A_{01}|^2A_{31}\nonumber\\
&
&-6A_{01}A_{21}A_{11}^*-6A_{23}A_{01}^*A_{11}^*-6A_{01}A_{11}A_{21}^*\nonumber\\
&
&-3A_{01}^2A_{31}^*-6\bar{\psi}A_{32}A_{01}^*-3A_{33}{A_{01}^*}^2\nonumber\\
& &-6\bar{\psi}A_{22}A_{11}^*+6i\partial_X(2A_{01}|A_{11}|^2+2|A_{01}|^2A_{21}\nonumber\\
& &+A_{01}^*A_{11}^2+A_{01}^2A_{21}^*)+3\partial_X^2(A_{01}^2A_{11}^*\nonumber\\
& &+2|A_{01}|^2A_{21}^*)+ \mbox{h.o.t.}\nonumber\\
\partial_TA_{41} &=&\partial_X^6A_{01}+6i\partial_X^5A_{11}-13\partial_X^4A_{21}-12i\partial_X^3A_{31}\nonumber\\
& &+4\partial_X^2A_{41}-(1-3\bar{\psi}^2)(2i\partial_XA_{31}+\partial_X^2A_{21})\nonumber\\
&
&+(1-3\bar{\psi}^2)A_{41}-3A_{21}^2A_{01}^*-6A_{11}A_{31}A_{01}^*\nonumber\\
& &-6A_{01}A_{41}A_{01}^*-6A_{11}A_{21}A_{11}^*-6A_{01}A_{31}A_{11}^*\nonumber\\
&
&-3A_{11}^2A_{21}^*-6A_{01}|A_{21}|^2-6A_{01}A_{11}A_{31}^*\nonumber\\
&
&-3A_{01}^2A_{41}^*-6\bar{\psi}A_{42}A_{01}^*-3A_{43}{A_{01}^*}^2\nonumber\\
&
&-6\bar{\psi}A_{32}A_{11}^*-6A_{33}A_{01}^*A_{11}^*-3A_{23}{A_{11}^*}^2\nonumber\\
& &-6\bar{\psi}A_{33}A_{22}^*-6A_{01}|A_{23}|^2+6i\partial_X(2A_{11}A_{21}A_{01}^*\nonumber\\
& &+2|A_{01}|^2A_{31}+A_{11}|A_{11}|^2+2A_{01}A_{21}A_{11}^*\nonumber\\
&
&+2A_{01}A_{11}A_{21}^*+A_{01}^2A_{31}^*)+3\partial_X^2(2A_{01}|A_{11}|^2\nonumber\\
& &+A_{01}^*A_{11}^2+2|A_{01}|^2A_{21}+A_{01}^2A_{21}^*) + \mbox{h.o.t.}
\end{eqnarray}
Here, ``$\mbox{h.o.t.}$'' refers to higher order terms that are functions of $A_{01}$ and its derivatives. The
amplitude function for the pattern ($e^{ix}$) can be written as
\begin{equation}
\label{eqn:msamppseries} A(X,T) = A_{01}(X,T) + \delta A_{11}(X,T) + \delta^2 A_{21}(X,T) + \ldots.
\end{equation}
Using Eqs.~(\ref{eqn:msamp3}), (\ref{eqn:msamp4to7}), and (\ref{eqn:msamppseries}), and scaling back to original
variables, i.e.~$X\to\delta^{-1}x$ and $T\to\delta^{-2}t$, the amplitude equation to $\mathcal{O}(\delta^4)$ can
be written as
\begin{eqnarray}
\partial_tA &=&
4\partial_x^2A-12i\partial_x^3A-13\partial_x^4A+6i\partial_x^5A+\partial_x^6A\nonumber\\
& &-\delta^2(1-3\bar{\psi}^2)\left(2i\partial_x+\partial_x^2\right)A +
\delta^2[(1-3\bar{\psi}^2)A\nonumber\\
& &-3A|A|^2+3\left(2i\partial_x+\partial_x^2\right)(A|A|^2)]-\delta^4(\frac{3}{64}A|A|^4\nonumber\\
& &+2\bar{\psi}^2A|A|^2)+\mathcal{O}(\delta^6)
\end{eqnarray}
or more compactly, after replacing $\delta^2\to\epsilon$, to $\mathcal{O}(\epsilon^2)$
\begin{eqnarray}
\label{eqn:ms1damp}
\partial_tA &=&
-\left(1-\mathcal{L}_{1D}\right){\mathcal{L}_{1D}}^2A-\epsilon(1-3\bar{\psi}^2)\mathcal{L}_{1D}A\nonumber\\
& &+\epsilon(1-3\bar{\psi}^2)A-3\epsilon
A|A|^2+3\epsilon\mathcal{L}_{1D}(A|A|^2)\nonumber\\
& &-\epsilon^2(\frac{3}{64}A|A|^4+2\bar{\psi}^2A|A|^2)+\mathcal{O}(\epsilon^3),
\end{eqnarray}
where $\mathcal{L}_{1D}\equiv(2i\partial_x+\partial_x^2)$.

Let us now compare Eq.~(\ref{eqn:ms1damp}) with the one dimensional
equivalents of the $\mathcal{O}(\epsilon)$ amplitude equations that we
have previously derived for hexagonal patterns,
i.e.~Eqs.~(\ref{eqn:qdamp1}) and (\ref{eqn:full_PRG_eps}). Without
re-deriving, the one dimensional equivalents are readily obtained by
setting
\begin{equation}
A_2=A_3=0\;\;\;\;\mbox{and}\;\;\;\;\mathcal{L}_{\mathbf{k}_2}=\mathcal{L}_{1D}
\end{equation}
in those equations. We observe that Eq.~(\ref{eqn:ms1damp}) (also
truncated to $\mathcal{O}(\epsilon)$) contains at least one term which
is not present in either of the equations previously derived.

We note that the QDRG result of Eq. (\ref{eqn:qdamp1}) is closer to the
multiple scales result compared to the proto-RG result (and the RG
result in section \ref{RG}), in that it fails to capture only the
nonlinear derivative term at $O(\epsilon)$, which is actually a higher
order correction to $3A|A|^2$. This is clearly because the spatial
operator in the QDRG method is an outcome of a \emph{linear} stability
analysis, whereas one would have to perform a nonlinear stability
analysis to obtain nonlinear spatial derivative terms. The clear
advantage of the QDRG calculation however is that it was done with
significantly less effort, and in a rotationally-covariant manner;
perturbation theory to $O(\epsilon)$ was all that was required. The
multiple scales analysis, on the other hand, required a sixth order
perturbation theory treatment, and in order to simplify the algebra, we
only worked in one dimension.  In higher dimensions, the interference
between the modes would have created a huge increase in the complexity
at each successively higher order in perturbation theory. The QDRG
calculation is only heuristic, but as we will show below, can be
justified from a full calculation, albeit with a minor technical
modification of the previously-published recipe, to take into account
the special feature of the conservation law in the PFC model.


We conclude therefore that although the QDRG result of
Eq.~(\ref{eqn:qdamp1}) and the multiple scales method to $O(\epsilon)$
still do not yield consistent results, the QDRG method is still an
improvement over the proto-RG method. In order to track down the source
of the discrepancy, we next attempted a full RG calculation without any
shortcuts, i.e. systematically calculating explicitly and renormalizing
all the divergent terms to $O(\epsilon)$.

\section{\label{RG}Renormalization group derivation}
In this section, we present a derivation of the amplitude equation
using the conventional RG method, in one dimension for pedagogical
simplicity (just as done for the method of multiple scales).  The
calculation is complicated because of the need to obtain explicit
formulae for the secular divergences\cite{CGO2,Graham}, but this is
possible at the order to which we worked.

Starting from Eq.~(\ref{eqn:pfc_scaled_1d}) (with $\delta^2$ replaced
by $\epsilon$) and a naive perturbation series in $\epsilon$ as in
Eq.~(\ref{eqn:pseries}) the zeroth and first order solutions can be
written as
\begin{eqnarray}
\psi_0 &=& \bar{\psi} + Ae^{ix} + \mbox{c.c.}\nonumber\\
\psi_1 &=& P_1(x,t)e^{ix}+Q_1e^{2ix}+R_1e^{3ix}+\mbox{c.c.}
\end{eqnarray}
The difficulty in the conventional RG method comes from the need to
explicitly determine the form of the secular coefficient $P_1$. While
this is a routine task for ODEs, it is far from trivial for PDEs. A
further complication is that the solution for $P_1$ must be the highest
order polynomial that satisfies the PDE, to be able to eliminate
\emph{all} secular divergences. It turns out that this is critical to
obtaining the rotationally covariant operator at a lower order in
$\epsilon$. Using the method of undetermined coefficients we find such
a solution to be
\begin{equation}
P_1(x,t) = \epsilon(1-3\bar{\psi}^2-3|A|^2)A\sum_{j=1}^6C_jP_{1j}(x,t),
\end{equation}
where
\begin{eqnarray}
P_{11} &=& t\nonumber\\
P_{12} &=&
-\frac{1}{720}(-89280t^2+7680t^3+4320ixt-34560ixt^2\nonumber\\
& &-4680x^2t+2880x^2t^2-1440ix^3t+120x^4t+x^6)\nonumber\\
P_{13} &=&
\frac{i}{720}(-5760it^2-1560xt+960xt^2-720ix^2t\nonumber\\
& &+80x^3t+x^5)\nonumber\\
P_{14} &=& \frac{1}{312}(192t^2-288ixt+48xt^2+x^4)\nonumber\\
P_{15} &=& -\frac{i}{72}(24xt+x^3)\nonumber\\
P_{16} &=& -\frac{x^2}{8},
\end{eqnarray}
and the constants $C_j$ satisfy $\sum_{j=1}^6C_j=1$.

The RG method proceeds as follows: (1) dummy variables $X$ and $T$ are
introduced, (2) the divergent terms in $P_{1j}$ of the form $x^mt^n$
are split to read $x^mt^n=(x^mt^n-X^mT^n) + X^mT^n$, (3) the constant
amplitude A is redefined using an $\epsilon$ expansion
$A=A_R(X,T)(1+\sum_{j=1}\epsilon^jZ_j)$, where $A_R$ is now the
renormalized amplitude, and $Z_j$ are the renormalization constants
which are chosen order by order in $\epsilon$ to absorb the $X^mT^n$
terms, and (4) since the solution $\psi$ is independent of $X$ and $T$,
all derivatives of $\psi$ with respect $X$, $T$, or a combination
thereof must be zero. This last condition yields the following RG
equations at $\mathcal{O}(\epsilon)$,
\begin{eqnarray}
\label{eqn:rgeqns} \frac{\partial A_R}{\partial T} &=&
C_1\epsilon(1-3\bar{\psi}^2-3|A|^2)A\nonumber\\
-\frac{\partial^6 A_R}{\partial X^6} &=& C_2\epsilon(1-3\bar{\psi}^2-3|A|^2)A\nonumber\\
-6i\frac{\partial^5 A_R}{\partial X^5} &=& C_3\epsilon(1-3\bar{\psi}^2-3|A|^2)A\nonumber\\
13\frac{\partial^4 A_R}{\partial X^4} &=& C_4\epsilon(1-3\bar{\psi}^2-3|A|^2)A\nonumber\\
12i\frac{\partial^3 A_R}{\partial X^3} &=& C_5\epsilon(1-3\bar{\psi}^2-3|A|^2)A\nonumber\\
-4\frac{\partial^2 A_R}{\partial X^2} &=& C_6\epsilon(1-3\bar{\psi}^2-3|A|^2)A.
\end{eqnarray}
Further, using $\sum_{j=1}^6C_j=1$ and replacing $A_R \rightarrow A$, $X \rightarrow x$, and $T \rightarrow t$,
the above equations can be combined to read
\begin{equation}
\label{eqn:rg1d}
\partial_tA + \left(1-\mathcal{L}_{1D}\right){\mathcal{L}_{1D}}^2A
=\epsilon(1-3\bar{\psi}^2)A-3\epsilon A|A|^2,
\end{equation}
which is also the 1-D proto-RG equation.

We close this section with some interesting observations. (i) The
equations in (\ref{eqn:rgeqns}) do not form a unique set of solvability
conditions. Other equations are possible, e.g.
\begin{eqnarray}
-\frac{1}{16}\frac{\partial^4 A_R}{\partial X^2 \partial T^2} &=& C_2\epsilon(1-3\bar{\psi}^2-3|A|^2)A\nonumber\\
-\frac{3i}{8}\frac{\partial^3 A_R}{\partial X \partial T^2} &=&
C_3\epsilon(1-3\bar{\psi}^2-3|A|^2)A \nonumber\\
&\vdots&
\end{eqnarray}
The choice of Eqn.~(\ref{eqn:rgeqns}) is motivated by the observation
that it yields a rotationally covariant amplitude equation, and other
physical considerations such as the microscopic equation being only
first order in time. (ii) The list of possible terms $P_{1j}$ does not
include the leading polynomial term $Bx$, where $B$ is an arbitrary
constant, as this term is annihilated by the kernel of the PDE. Thus no
constraint is available to fix $B$. It turns out that unless this term
is also renormalized, all secular divergences are not removed. This may
explain the absence of certain terms in Eq.~(\ref{eqn:rg1d}) that
however show up in the multiple scales analysis. To be certain, the
calculation needs to be carried out to higher orders;  but we do not
attempt this here.

\section{\label{MODI}Operator ordering ambiguity and its resolution in the RG method}

In this section, we resolve the discrepancy between the answers
generated by the QDRG method, the RG methods, and the method of
multiple scales. Curiously, no such discrepancy was observed in the
treatment of the Swift-Hohenberg equation, a non-conservative OPE, by
RG methods \cite{CGO2,Graham,Nozaki01} and multiple scales techniques
\cite{gunaratne}. In fact, it can also be readily ascertained that the
QDRG method will produce the same result as the other methods for this
equation, which we leave as a simple exercise for the reader. Why then
does a discrepancy arise in the PFC equation? Clearly, the role played
by the extra Laplacian, a consequence of the conservation law in this
case, must be non-trivial!

Note that this Laplacian operator carries over to the right hand side
of both Eq.~(\ref{eqn:PRG1}), the $\mathcal{O}(\epsilon)$ equation for
the RG methods, and Eq.~(\ref{eqn:ms3}), the $\mathcal{O}(\epsilon)$
equation for multiple scales. However also note that, in the method of
multiple scales, in addition to the non-linear terms, this operator is
also subjected to an $\epsilon$ expansion. There is no provision in
\emph{any} of the RG methods to allow the same to happen to the
Laplacian. In other words, the operator may very well have not existed
on the right hand side at $\mathcal{O}(\epsilon)$, and we would have
obtained exactly the same result as before!

A clue to the subtlety is to look at the way in which the secular terms
are renormalized. The naive way, as followed here, would be to evaluate
the right hand side first, look for secular terms later, and then
renormalize these divergent coefficients. However, this will not
eliminate secular terms generated by the differential operator. In
order to eliminate \emph{all} secular terms, the amplitude must be
renormalized \emph{before} differentiation, for the simple reason that
renormalization and differentiation are non-commutable operations. In
other words, there is an operator ordering ambiguity in the
implementation of the renormalization group method, exposed in this
problem by the conservation law.  Performing the calculation with the
operations of renormalization and differentiation reversed is
equivalent to performing an $\epsilon$ expansion in the differential
operator.

We find that by following this procedure, additional terms in the
coefficients of the resonant modes are automatically generated.
Specifically, when we evaluate the right hand side of
Eq.~(\ref{eqn:PRG1}) after assuming the amplitudes of $\psi_0$ to have
a space-time dependence, the renormalized coefficients of the resonant
$\exp(i\mathbf{k}_1\cdot\mathbf{x})$ forcing term work out to be
\begin{eqnarray}
\label{eqn:nlnrcorr} & &\epsilon[(1-3\bar{\psi}^2)A_{1}-3
A_{1}\left(|A_{1}|^2+2|A_{2}|^2+2|A_{3}|^2\right)\nonumber\\
&
&-6A_{2}^*A_{3}^*\bar{\psi}-(1-3\bar{\psi}^2)\mathcal{L}_{\mathbf{k}_1}A_1\nonumber\\
& &+6\left(|A_{1}|^2+|A_{2}|^2+|A_{3}|^2\right)\mathcal{L}_{\mathbf{k}_1}A_1+6A_1^*|\nabla
A_1|^2\nonumber\\
&
&+3A_1^2\mathcal{L}_{\mathbf{k}_1}A_1^*+6A_1A_2^*\mathcal{L}_{\mathbf{k}_1}A_2+6A_1A_2\mathcal{L}_{\mathbf{k}_1}A_2^*\nonumber\\
&
&+6A_1A_3^*\mathcal{L}_{\mathbf{k}_1}A_3+6A_1A_3\mathcal{L}_{\mathbf{k}_1}A_3^*+6\bar{\psi}A_3^*\mathcal{L}_{\mathbf{k}_1}A_2^*\nonumber\\
&
&+6\bar{\psi}A_2^*\mathcal{L}_{\mathbf{k}_1}A_3^*+12\bar{\psi}\nabla A_2^*\cdot\nabla A_3^*\nonumber\\
& &+12A_1\left(\nabla A_1\cdot\nabla A_1^*+\nabla
A_2\cdot\nabla A_2^*+\nabla A_3\cdot\nabla A_3^*\right)\nonumber\\
& &+12A_2\nabla A_1\cdot\nabla A_2^*+12A_2^*\nabla A_1\cdot\nabla A_2\nonumber\\
& &+12A_3\nabla A_1\cdot\nabla A_3^*+12A_3^*\nabla A_1\cdot\nabla A_3],
\end{eqnarray}
which when specialized for the 1-D case becomes
\begin{eqnarray}
&
&\epsilon[(1-3\bar{\psi}^2)A-3A|A|^2-(1-3\bar{\psi}^2)\mathcal{L}_{1D}A\nonumber\\
&&+6|A|^2\mathcal{L}_{1D}A+3A^2\mathcal{L}_{1D}A^*+6A^*\left(\frac{\partial
A}{\partial x}\right)^2\nonumber\\
&&+12A\frac{\partial A}{\partial x}\frac{\partial A^*}{\partial
x}]\nonumber\\
&=&\epsilon[(1-3\bar{\psi}^2)A-3A|A|^2-(1-3\bar{\psi}^2)\mathcal{L}_{1D}A\nonumber\\
&&+3\mathcal{L}_{1D}(A|A|^2)].
\end{eqnarray}
We note that the above terms are identical to the
$\mathcal{O}(\epsilon)$ terms on the right hand side of
Eq.~(\ref{eqn:ms1damp}). Therefore the correct amplitude equation to
$\mathcal{O}(\epsilon)$ should contain all the terms in
Eq.~(\ref{eqn:nlnrcorr}). In order to illustrate the generality of this
approach, we apply this idea again in the appendix to the Van der Pol
oscillator, another equation for which the previously reported
implementation of the RG method, and the method of multiple scales
produce different answers.

We wish to point out that the assumption of a constant amplitude in the
$\psi_0$ solution makes it possible for the coefficients of the
non-resonant terms in $\psi_1$ to assume constant values, a fact that
is favorably used in extending the proto-RG calculation to the next
order. However, with our modification to the proto-RG procedure, it is
clear that for the PFC equation at least, non-resonant coefficients
cannot have constant values. Thus computing higher order corrections to
the amplitude equations, will require explicit construction of
particular solutions, which may limit progress beyond
$\mathcal{O}(\epsilon)$ by purely analytical methods.

\section{\label{CONC}Conclusion}

We have presented a detailed illustration of various perturbative
techniques to derive amplitude equations from order parameter equations
that produce periodic patterns. Amplitude equations serve as powerful
analytical tools with which to investigate pattern stability and defect
interactions, as well as accurate coarse-grained descriptions of
pattern forming systems, and this calls for practical and reliable
mathematical methods for deriving them.

Although our benchmark for accuracy is the widely-accepted method of
multiple scales, it is critical to note that this method is not
failsafe, because it requires {\it a priori\/} identification of the
way in which space and time scale with the small parameter $\epsilon$.
There are many instances where surprising scales emerge that would not
easily be identified {\it a priori\/} (e.g. see the analysis of the
Mathieu equation in \cite{CGO2}).

The method of multiple scales typically involves a very lengthy
calculation before a rotationally covariant operator ensues, and
involves computation of various higher order terms which ultimately do
not improve the overall result significantly.  In the example presented
here, a sixth order calculation was required to get the lowest order
amplitude equation.  The reader should bear in mind that the fairly
involved calculation shown in this article was only one dimensional.

On the other hand, the practicality of RG based methods, where the
amplitude equation was obtained very quickly at $\mathcal{O}(\epsilon)$
itself, is self-evident. No guesswork was required to determine the
scaling of the variables and all calculations started with naive
perturbation expansions in $\epsilon$. In particular, our so called
``quick and dirty'' (QDRG) method and the proto-RG method are
attractive techniques, because there is virtually no need to construct
explicit solutions. Both methods use only information available from
the differential equation and in that sense, are very general ways of
building a controlled coarse-grained approximation to the order
parameter equation being studied. Furthermore, the QDRG method gives
the correct result quickly, apart from a small non-linear rotationally
covariant gradient term which is not captured by the linear stability
argument.

At $\mathcal{O}(\epsilon)$, we have shown that the QDRG method produces
a more accurate amplitude equation compared to the proto-RG method, by
capturing certain extra terms that are revealed in the multiple scales
analysis. However, with our corrected order of operators in the way in
which the RG is implemented, we find that all methods converge
identically at this order.

In conclusion, we have presented a detailed calculation of the
coarse-graining of the phase field crystal equation, for small
$\epsilon$.  Elsewhere \cite{GAD05_1, GAD05_2}, we have demonstrated
the utility of the coarse-grained equation in performing large-scale
simulations of materials processing phenomena in two dimensions.
Further developments of these techniques are underway, implementing
adaptive mesh refinement to solve the amplitude equations obtained here
in two and three dimensions, and going beyond the single mode
approximation which has formed the basis of our work and that of Elder
and collaborators.  We hope to report on these developments at a future
date.

\begin{acknowledgments}

This work was partially supported by the National Science Foundation
through grant number NSF-DMR-01-21695. The authors would like to thank
Prof.~Sasha Golovin for useful discussions and bringing reference
\cite{MATT00} to our attention.

\end{acknowledgments}

\appendix*

\section{Van der Pol Oscillator}

In this appendix, we explore the commuting of differentiation and
renormalization with a simple ordinary differential equation example:
the Van der Pol oscillator. Note that this yet another case in which a
differential operator is multiplied by a small parameter (see right
hand side of Eq.(\ref{eqn:vpol})).

The autonomous ODE is given by
\begin{equation}\label{eqn:vpol}
y'' + y = \epsilon(1-y^2)y',
\end{equation}
where $'$ denotes differentiation with respect to the variable $t$. As
there is a derivative on the right hand side of this equation we
anticipate that the proto-RG amplitude equation will fail to capture
certain terms that turn out in the multiple scales analysis.

It is known that the scaling $\tau = \epsilon t$ works for this problem \cite{Orszagbook}. Hence,
\begin{eqnarray}
y' &\rightarrow& (\partial_t + \epsilon \partial_{\tau})y\nonumber\\
y'' &\rightarrow& (\partial_t^2 + 2\epsilon
\partial_{\tau}\partial_t + \epsilon^2 \partial_{\tau}^2)y,
\end{eqnarray}
where the subscripts denote partial differentiation. Expanding $y$ in a perturbation series
\begin{equation}
y = y_0 + \epsilon y_1 + \epsilon^2 y_2 + \ldots
\end{equation}
we obtain
\begin{eqnarray}
\mathcal{O}(1): (\partial_t^2+1)y_0 &=& 0\nonumber\\
\mathcal{O}(\epsilon): (\partial_t^2+1)y_1 &=&
-2\partial_{\tau}\partial_ty_0 + (1-y_0^2)\partial_ty_0\nonumber\\
\mathcal{O}(\epsilon^2): (\partial_t^2+1)y_2 &=& -2\partial_{\tau}\partial_ty_1 + (1-y_0^2)\partial_ty_1-\partial_{\tau}^2y_0\nonumber\\
& &-2y_0y_1\partial_ty_0+(1-y_0^2)\partial_{\tau}y_0.
\end{eqnarray}
From this we find
\begin{eqnarray}
y_0 &=& A_{01}(\tau)e^{it} + \mbox{c.c.}\nonumber\\
y_1 &=& A_{11}(\tau)e^{it} + A_{13}(\tau)e^{3it} + \mbox{c.c.}
\end{eqnarray}
Application of the Fredholm alternative at $\mathcal{O}(\epsilon)$ and $\mathcal{O}(\epsilon^2)$ yields the
following amplitude equations
\begin{eqnarray}
2i\partial_{\tau}A_{01} &=& iA_{01}\left(1-|A_{01}|^2\right)\nonumber\\
\partial_{\tau}^2A_{01}+2i\partial_{\tau}A_{11}&=&i\left(A_{11}-2A_{11}|A_{01}|^2-A_{01}^2A_{11}^*\right)\nonumber\\
& &+\partial_{\tau}\left(A_{01}-A_{01}|A_{01}|^2\right)\nonumber\\
& &+\frac{A_{01}|A_{01}|^4}{8}
\end{eqnarray}
which can be combined after scaling back to original variables to get
\begin{equation}
\label{eqn:appms}
\partial_t^2A + 2i\partial_tA =
\epsilon\left[iA(1-|A|^2)+\partial_tA(1-|A|^2)\right] + \mathcal{O}(\epsilon^2).
\end{equation}

Nozaki and Oono \cite{Nozaki01} on the other hand have obtained the following equation using the proto-RG method
\begin{equation}
\partial_t^2A + 2i\partial_tA =
\epsilon i A(1-|A|^2) + \mathcal{O}(\epsilon^2).
\end{equation}
Note that the missing term $\partial_tA(1-|A|^2)$ can be captured by differentiating the lower order result,
i.~e.
\begin{equation}
2i\partial_tA = \epsilon i A(1-|A|^2)
\end{equation}
but this does not seem a very general approach. In particular, it is not obvious how this can be extended to
PDEs.

The $\mathcal{O}(\epsilon)$ equation using the proto-RG method reads
\begin{equation}
y_1''+y_1 = (1-y_0^2)y_0',
\end{equation}
where
\begin{eqnarray}
y_0 &=& A e^{it} + \mbox{c.c.}\nonumber\\
y_1 &=& P(t) e^{it} + Q e^{3it} + \mbox{c.c.}
\end{eqnarray}
where $A$ can be a constant while $P$ cannot. Thus, the proto-RG operator turns out to be
\begin{equation}
\mathcal{L} = \partial_t^2 + 2i\partial_t,
\end{equation}
and the proto-RG equation reads
\begin{equation}
\label{eqn:app_prg} \mathcal{L}A = \epsilon\mathcal{L}P + \mathcal{O}(\epsilon^2),
\end{equation}
where A is now the renormalized amplitude. When evaluating
$\mathcal{L}P$ however, we allow for the possibility that $A$, which
appears on the right hand side of the equation can also be a function
of $t$, or equivalently renormalize $A$ on the right hand side before
differentiating $y_0$, which gives us
\begin{equation}
\mathcal{L}P = \epsilon\left[iA(1-|A|^2)+\partial_tA(1-|A|^2)\right].
\end{equation}
Therefore the true amplitude equation should read (using Eq.~\ref{eqn:app_prg})
\begin{equation}
\mathcal{L}A = \epsilon\left[iA(1-|A|^2)+\partial_tA(1-|A|^2)\right] + \mathcal{O}(\epsilon^2),
\end{equation}
which is identical to the multiple scales result of Eq.~(\ref{eqn:appms}).


\bibliographystyle{apsrev}

\end{document}